\documentclass[a4paper,11pt]{article}
\pdfoutput=1 

\usepackage{jheppub} 

\usepackage[T1]{fontenc} 
\usepackage{tikz}

\usepackage{multirow}

\title{Interacting CFTs for all couplings: Thermal versus Entanglement Entropy at Large $N$}

\author[a]{S. Grable}

\affiliation[a]{University of Colorado Boulder,\\2000 Colorado Ave, Boulder, CO 80309 United States of America}

\emailAdd{seth.grable@colorado.edu}

\abstract{In this paper, I calculate the large $N$ limit of marginal $O(N)$ models with non-polynomial potentials in arbitrary odd dimensions $d$. This results in a new class of interacting pure conformal field theories (CFTs) in \(d=3+4n\) for any $n \in \mathbb{Z}_+$. Similarly, in  \(d=3+4n\) I calculate the thermal entropy for all couplings on $R^{2+4n} \times S^1$ for \(n=0,1,2,3\). In 2+1 dimensions I find the strong-to-weak coupling ratio of the thermal entropy to be 4/5, matching recent results, and further extend this analysis to higher odd dimensions. Next, I calculated the vacuum entanglement entropy $s^d_{\text{EE}}$ on \(S^{d-2}\) for all couplings in arbitrary odd $d$ in the large N limit. I find the vacuum entanglement entropy on \(S^{d-2}\) to be not only solvable but also constant for all couplings $\lambda$. Thus, in the large $N$ limit, the vacuum entanglement entropy on \(S^{d-2}\) for odd \(d\) is constant for all $\lambda$, in contrast to the thermal entropy which is shown to also be monotonically decreasing with $\lambda$ in \(d=3+4n\).}

\begin{document} 
\maketitle
\flushbottom

\section{Introduction}
\label{sec:intro}
Pure CFTs are a sub-class of conformal field theories (CFTs). CFTs  main characteristic is a vanishing beta-function and can be found at critical parameter values within a larger parent theory. However, pure CFTs maintain conformal invariance for all values of a parameter, such as the coupling \(\lambda\). Further, some CFTs in the large N-limit give exact analytic results for all \(\lambda\). A well know example of pure CFT is a \(\mathcal{N} = 4\) Super-Yang-Mills theory in 3+1 dimensions whose free energy at finite temperature and infinite coupling, via its conjectured gravity dual \cite{maldacena}, is 3/4 of the Stefan-Boltzmann free theory \cite{gubser}\cite{dewolfe}. However, the entropy density \(s\) of the \(\mathcal{N}\) = 4 SYM theory in four dimensions is not solvable for all \(\lambda\) \cite{romatschke}\cite{blaizot}. Nonetheless, in the large N limit, I find a large class of field theories, in \(d=3+4n\) for any $n \in \mathbb{Z}_+$, that are exactly solvable for all couplings. It is no secret that purely solvable models for all couplings are rare, and as with quantum mechanics more broadly, analytically solvable models often bring new insight. 

 In the large N limit Ref \cite{romatschke} shows a \(\phi^4\) finite-temperature \(O(N)\) model in \(2 + 1\) dimensions on flat space is solvable for all \(\lambda\), and has a weak-to-strong entropy density ratio of \(4/5\). This result was then extended to large \(N\) supersymmetric Wess-Zumino model in \(2+1\) dimension where the weak-to-strong entropy density ratio was found to be \(\frac{31}{35}\) \cite{dewolfe}.
\\\indent{} In this spirit I continue to investigate large \(N\) pure CFT's in higher odd dimensions. I start by presenting a generalized marginal vector model in arbitrary odd dimensions. All theories under this generalization lack logarithmic divergences, and theories in \(d=3+4n\) yield gap equations with non-trivial finite solutions. This results in a new class of interacting pure CFTs. Further, applying this generalized theory to \(2+1d\) produces a strong-to-weak thermal entropy density ratio of \(\frac{4}{5}\), matching the results of quartic and sextic theories in Refs. \cite{romatschke}\cite{romatschkeshear} and \cite{pinto}. 

In Refs. \cite{casini} and \cite{Myers} it was shown that the vacuum entanglement entropy on a spherical region in \(R^{d-1}\times S^1\) maps to the thermal entropy in \(S^d\). Further, it was shown that the non-universal piece of the vacuum entanglement entropy on the boundary of a spherical region is positively divergent. Removing the non-universal piece via regularization leaves a finite universal piece which can be either positive or negative depending on dimensionality. With this, I show in the large \(N\) limit that the entanglement entropy on \(S^{d-2}\) defined as \(s^d_{\text{EE}}\), is constant for all \(d\), further implying the free propagator is vanishing in all such theories. Finally, \(s^d_{\text{EE}}\) in the large N limit is shown to match \(N\) copies of the thermal entropy of a free theory on \(S^d\) for any odd \(d\), given by Ref. \cite{klebanov}.

The contrast between results of various geometries prompts further discussion on large \(N\) theories on curved spacetime, particularly on how the d-sphere acts as a local minimum of entanglement entropy among all shapes \cite{mezei}, and the possibility of conformally deforming the minimized entanglement entropy \(S^d_{\text{EE}}\) to other topologies in the large \(N\) limit. Further, investigation can be extended to Fermionic large \(N\) theories, and to fully solvable higher-spin gravity duals of large \(N\) theories. 

\section{The Model}
\label{sec:model}
\subsection{A Marginal Theory on Flat Space in Odd Dimensions}
\label{sec:flat space}
Consider a marginal \(O(N)\) bosonic scalar theory in odd dimensions where \(\vec{\phi} = (\phi_1, \phi_2 ... \phi_N\)) in the limit \(N\rightarrow \infty\),  with the Euclidean action 
\begin{equation}\label{eq1}
 A^{(d)} =\frac{1}{2} \int_{x} \bigg[\partial_\mu \vec{\phi}\partial_\mu \vec{\phi} + m_0^2\vec{\phi}^2 + \bigg(\frac{\lambda}{N}\bigg)^\frac{2}{d-2}\big(\vec{\phi}\cdot\vec{\phi}\big)^{\frac{d}{d-2}}\bigg].
\end{equation}
As standard practice, let the Euclidean dimension of time be compactified to a thermal circle of radius \(\beta\) \cite{laine}. Next the partition function \(Z^{(d)}(T) = \int \mathcal{D} \vec{\phi} e^{-A^{(d)}}\) is multiplied by the identity \cite{romatschke} 
\begin{equation}\label{eq2}
    1 = \int \mathcal{D}\xi\mathcal{D}\sigma e^{i\int_x \frac{1}{2}\xi (\sigma - \frac{\vec{\phi}\cdot\vec{\phi}}{N})},  
\end{equation}
 and for simplicity bare mass term \(m_0\) is tuned to zero. Further let \( \xi \rightarrow N\xi\) and \(\sigma \rightarrow \frac{\sigma}{\lambda}\) to evenly distribute the coupling in \(Z(T)\). Next saddle point integrals are taken around the mean-field values of \(\xi\) and \(\sigma\), \(\Bar{\xi}\) and \(\Bar{\sigma}\), which are exact in the large N limit. Applying the saddle point condition for the \(\sigma\) field I keep only the positive solution in terms of \(\Bar{\xi}\), which is equivalent to keeping only positive effective mass terms. Now \(Z(T)\) is
\begin{equation}\label{eq3}
\begin{split}
 Z^{(d)}(T) = \int \mathcal{D}\phi d\Bar{\xi} \exp \bigg[-N\int_x \Bigg( \frac{1}{2}\phi \big( \partial_\mu \partial_\mu + i\Bar{\xi} \big)\phi - \frac{(i\Bar{\xi})^{d/2}}{ \lambda}\bigg(\frac{2}{d}\bigg)\bigg(\frac{d-2}{d}\bigg)^{\frac{d-2}{2}}\Bigg)\bigg],
\end{split}
\end{equation}
giving a marginally coupled theory for any \(d\), however, this paper is specifically investigating theories in odd dimensions. Letting \(\Bar{i\xi}\rightarrow m^2\) in noting \(\Bar{i\xi}\) acts as the effective mass, up to an overall constant the partition function is
\begin{equation}\label{eq4}
 Z^{(d)}(T)=\exp\bigg[  N\beta V\Bigg( P^{(d)}_{\text{free}}\big(T,m\big) +  \frac{m^{d}}{ \lambda}\bigg(\frac{2}{d}\bigg)\bigg(\frac{d-2}{d}\bigg)^{\frac{d-2}{2}}\Bigg)\bigg],
\end{equation}
where \(P^{(d)}_{\text{free}} \big(T,m\big)\), the pressure of a free scalar field theory at temperature \(T\) in \(d\) dimensions, is given by 
\begin{equation}\label{eq5}
  P^{(d)}_{\text{free}}(T,m)= -\int \frac{d^{d-1}k}{(2\pi)^{d-1}}\bigg[\frac{E_k}{2} + T\ln (1 -e^{-\beta E_k})\bigg]. 
\end{equation}
Using dimensional regularization \(P^{(d)}_{\text{free}}\) evaluates to an exact analytic result of the form:
\begin{equation}\label{eq34}
    P^{(d)}_{\text{free}} = -\frac{1}{2}\frac{m^{d} \Gamma(-\frac{d}{2})}{(4\pi)^{\frac{d-1}{2}} \Gamma(-\frac{1}{2})} +\frac{T\Omega_{d-1}}{(2\pi)^{d-1}} \sum_{n=1}^{\infty}\int_{m}^{\infty} \frac{1}{n}(k^2-m^2)^{\frac{d-3}{2}}k e^{-n\beta k}dk,
\end{equation}
where \(\Omega_{d} = \frac{2\pi ^{d/2}}{\Gamma(\frac{d}{2})}\) is the solid angle, and \(d\) still holds as the dimensionality of theory. The integral in equation \eqref{eq34} can be evaluated for any \(d\) and \(m>0\) to give, 
\begin{equation}\label{eq29}
    P^{(d)}_{\text{free}} =\frac{m^d \Gamma(\frac{-d}{2})}{2^{d+1}\pi^{\frac{d}{2}}} +\frac{2^{1-\frac{d}{2}} m^d} {(\pi m \beta)^{\frac{d}{2}}}
   \sum_{n=1}^{\infty} \big(\frac{1}{n}\big)^{\frac{d}{2}} K_{\frac{d}{2}}\Big(m n \beta\Big),
\end{equation}
One can see that the dimensionally regulated piece of \eqref{eq29} has no logarithmic divergence in odd dimensions. To apply \eqref{eq29} consider \(d=2+1\) resulting in a \(\phi^6\) theory with marginal coupling, the free pressure evaluates to
\begin{equation}\label{eq6}
  P^{(3)}_{\text{free}} =  \frac{T^3}{2\pi} \big[\text{Li}_3(e^{-\beta m}) + \frac{m}{T}\text{Li}_2(e^{-\beta m}) + \frac{m^3}{6T^3} \big].
\end{equation}
The saddle condition of the \(\xi\)-field is given in terms of the effective mass as 
\begin{equation}\label{eq7}
\frac{2m}{\sqrt{3}\lambda}  = \frac{m}{4\pi} + \frac{1}{2\pi}T\ln (1 -e^{-\beta m}).  
\end{equation}
Applying the saddle point condition \eqref{eq7} to \eqref{eq4} gives 
\begin{equation}\label{eq8}
       P^{(3)}(T,\lambda) = N\bigg[\frac{T^3}{2\pi} \big[\text{Li}_3(e^{-\beta m}) + \frac{m}{T}\text{Li}_2(e^{-\beta m}) -\frac{m^2}{3 T^2}\ln (1- e^{-\beta m})\big]\bigg].
\end{equation}

The thermal entropy density is found via \(s_{\text{therm}
}= \frac{\partial P}{\partial T}\), resulting in
\begin{equation}\label{eq10}
    s^{(3)}_{\text{therm}}(T,\lambda) =  N\bigg[\frac{3T^2}{2\pi} \big[\text{Li}_3(e^{-m\beta}) +\frac{m}{T}  \text{Li}_2(e^{-m\beta} ) - \frac{m^2}{3T^2}\ln (1- e^{-m\beta })\big] \bigg].
\end{equation}
\(s^{(3)}_{\text{therm}}(T,\lambda)\) matches the bosonic contribution of the entropy density of a Nambu–Jona-Lasinio–Yukawa model given in Ref. \cite{pinto}. Considering equation \eqref{eq7}, the gap equation at infinite coupling is
\begin{equation}\label{eq11}
    -\beta m =2 \ln(1 - e^{-\beta m}). 
\end{equation}
This gap equation is solved in terms of the golden ratio \(\phi= \frac{1+\sqrt{5}}{2}\) giving
\begin{equation}\label{eq12}
   \beta m = 2\ln (\phi),
\end{equation}
which matches the infinite coupling gap equation of Ref. \cite{romatschke}. In the zero coupling limit, we have \(\beta m = 0\). Then evaluating \eqref{eq10} with \eqref{eq12} and \(\beta m = 0\), \(s^{(3)}_{\text{therm}}(T,0)\) and \(s^{(3)}_{\text{therm}}(T,\infty)\) are given as
\begin{equation}\label{eq14}
\begin{split}
    &s^{(3)}_{\text{therm}}(T,0) = \frac{3NT^2\zeta(3)}{2\pi} = s_{\text{therm}}^{\text{free}^{(3)}},\\&
    s^{(3)}_{\text{therm}}(T,\infty) =  \frac{6NT^2\zeta(3)}{5\pi} = s_{\text{therm}}^{\infty^{(3)}}.
\end{split}    
\end{equation}
These coupling limits match Ref. \cite{romatschke}\cite{pinto}, and \cite{romatschkeshear}, and the strong coupling limit was recognized earlier in Refs. \cite{sachdev}\cite{drummond}. Together the equations of \eqref{eq14} give a strong-to-weak coupling ration of,
\begin{equation}\label{eq15}
 \frac{s_{\text{therm}}^{\infty^{(3)}}}{s_{\text{therm}}^{\text{free}^{(3)}}} = \frac{4}{5}.
\end{equation}
This strong-to-weak entropy density ratio matches the results of a \(\phi^4\) theory in \(R^2\times S^1\) and has further been shown to be universal in a large class of pure CFTs with only bosonic degrees of freedom in \(R^2 \times S^1\) \cite{romatschke}. 

\subsection{Higher Dimensions on Flat Space}

\begin{figure}[h]
\centering
\includegraphics[width=.9\textwidth]{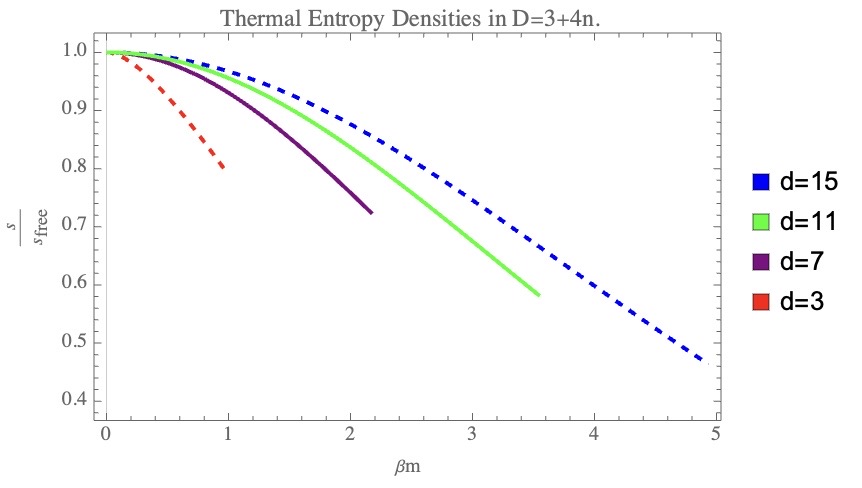}
\caption{\label{graph}The thermal entropy density of \(s_{\text{therm}}/s^{\text{free}}_{\text{therm}}\) are shown above for all temperatures/couplings ranging in order from \(d=3\) at the bottom to \(d=15\). }
\label{fig: plot}
\end{figure}

To find the gap equations for higher dimensions I will start at \(d=3\) and move up in increments of 4 dimensions. We can call this ``the rule of fours.'' The reasoning for this strategy goes as follows: in the \(\lambda\rightarrow\infty\) limit the finite temperature piece of equation \eqref{eq29} gives a series of polylogarithms of the form \(\text{Li}_n(e^{-m\beta})\). Now, for all finite \(m\), \(n\), and \(d\):
\begin{equation}
\begin{split}
    &\frac{\partial}{\partial m^2} \text{Li}_n(e^{-m\beta}) = -\text{Li}_{n-1}(e^{-m\beta}) < 0,
\end{split} 
\end{equation}
and,
\begin{equation}
    \lim_{m\to\infty}-\text{Li}_{n-1}(e^{-m\beta})=0.
\end{equation}
Further, the mass derivative of the zero temperature piece of the equation \eqref{eq29} is proportional to \(\Gamma(-\frac{d}{2})\)
\begin{equation}
    \frac{\partial}{\partial m^2} \frac{m^d \Gamma(-\frac{d}{2})}{2^{d+1}\pi^{\frac{d}{2}}}\sim \Gamma\bigg(-\frac{d}{2}\bigg).
\end{equation}
Finally, \(\Gamma\big(-\frac{d}{2}\big)>0\)  for \(d=3,7,11,15...\), for all \(d=3+4n\) with \(n\in \mathbb{N}\). Thus, the gap equations in dimensions \(d=3+4n\) are satisfied by some non-trivial finite mass value for any \(n\in \mathbb{N}\) at any finite temperature \(T\). This analysis can be extended to finite \(\lambda\) to include the coupling term in equation \eqref{eq4} which is overall positive for \(d>2\). For example, the gap equation in the limit \(\lambda\rightarrow\infty\) for \(d=7\) is given as
\begin{equation}
    \beta ^5 m^5=30 \beta ^2 m^2 \text{Li}_3\left(e^{-m \beta }\right)+90 \beta  m \text{Li}_4\left(e^{-m \beta }\right)+90 \text{Li}_5\left(e^{-m \beta }\right),
\end{equation}
which is numerically solved around \(\beta m \approx 2.178\). Likewise, the gap equation in the zero temperature limit is
\begin{equation}
\left(\frac{25 \sqrt{\frac{5}{7}}}{49 \lambda }+\frac{1}{480 \pi ^3}\right) m^5=0
\end{equation}
which satisfied at \(m=0\) for all \(\lambda>0\). Given that the gap equations for dimensions \(d=3+4n\) can be solved at finite temperature for all couplings \(\lambda\), and the corresponding theories contain no logarithmic divergences; in the large \(N\) limit, marginal theories in \(d=3+4n\) generate a large class of interacting pure CFTs. Figure (\ref{fig: plot}) shows \(P^{(d)}(T,m)\) for \(d=3,7,11,15\). Likewise the chart below gives \(s^{\text{free}}_{\text{therm}}\) and \(s_{\text{therm}}^{\infty}/{s_{\text{therm}}^{\text{free}}}\) values for \(d=3,7,11,15\).
\begin{center}
\begin{tabular}{ |p{3cm}||p{3cm}|p{3cm}| }
 \hline
 \multicolumn{3}{|c|}{Thermal Entropy Relations} \\
 \hline
 Dimensions& \(s^{\text{free}}_{\text{therm}}\) & \(s_{\text{therm}}^{\infty}/{s_{\text{therm}}^{\text{free}}}\)
 \\
 \hline
 \(d=3\) & \( \frac{3NT^2\zeta(3)}{2\pi}\) & = 0.800\\
 \(d=7\) &\(\frac{15NT^7\zeta(7)}{8\pi^3}\) & \(\approx 0.649\)\\
 \(d=11\)& \(\frac{945T^{11} \zeta (11)}{32 \pi ^5}\) &\(\approx 0.578\)\\
 \(d=15\)& \(\frac{135135 T^{15} \zeta (15)}{128 \pi ^7}\)  & \(\approx 0.466\)  \\
 \hline
\end{tabular}
\end{center}

However, \(\Gamma\big(-\frac{d}{2}\big)<0\) for all \(d=1+4n\) with \(n\in \mathbb{N}\). Thus, gap equations for dimensions \(d=1+4n\) in the \(\lambda\rightarrow\infty\) limit are overall negative, giving only trivial solutions at finite temperature. Consider the \(\lambda\rightarrow\infty\) limit for \(d=5\), the gap equation is given as

\begin{equation}
\frac{m^3\beta^3}{6}+\beta  m \text{Li}_2\left(e^{-m \beta}\right)+\text{Li}_3\left(e^{-m \beta }\right) = 0.
\end{equation}

 By physical demands \(m>0\) and \(\beta>0\) and because
\(\text{Li}_2\left (e^{-m \beta}\right)>0\) this gap equation is only trivially solved by \(m=0\) as \(\beta\rightarrow\infty\). We can also note that for \(d=1+4n\) in the zero temperature limit there is a critical coupling \(\lambda_{\text{crit}}\) which satisfies the saddle condition for all in medium mass giving a constant zero pressure for all \(m\). For \(d=5\) in the \(T\rightarrow 0\) limit, the saddle condition is
\begin{equation}
m^3 \left(\frac{3 \sqrt{\frac{3}{5}}}{5 \lambda }-\frac{1}{48 \pi ^2}\right)=0, 
\end{equation}
which is satisfied for all \(m\) at \(\lambda_{\text{crit}}= \frac{144}{5} \sqrt{\frac{3}{5}} \pi ^2\) giving \(P(0,m,\lambda_{\text{crit}})=0\; \forall\; m \). The physical implications of this critical phenomenon can be explored in future work.

\section{Entanglement Entropy on a d-Sphere}
\begin{figure}
    \centering
    \caption{The entanglement entropy of a spherical region in \(S^{d-1}\) embedded in \(R^{d-1}\times S^1\) is found on the boundary in \(S^{d-2}\).}
    \label{fig:my_label}
\begin{center}
\begin{tikzpicture}
\shadedraw[black, opacity = .4][ultra thick](-6,-3)-- (-4,3)--(6,3)--(4,-3)--(-6,-3);
\draw[blue][ultra thick](0,0) ellipse (1in and .75in);
\shade[ball color = cyan, opacity = .4] (0,0) ellipse (1in and .75in);
\draw node at (-3,2) {$R^{d-1}\times S^1$};
\draw node at (.5,0) {$S^{d-1}$};
\draw node at (3.5,-1) {$S^{d-2}$};
\draw[->][thick] (3.5,-.75)--(2.6,.25);
\end{tikzpicture}
\end{center}
\label{fig: entanglement entropy}
\end{figure}
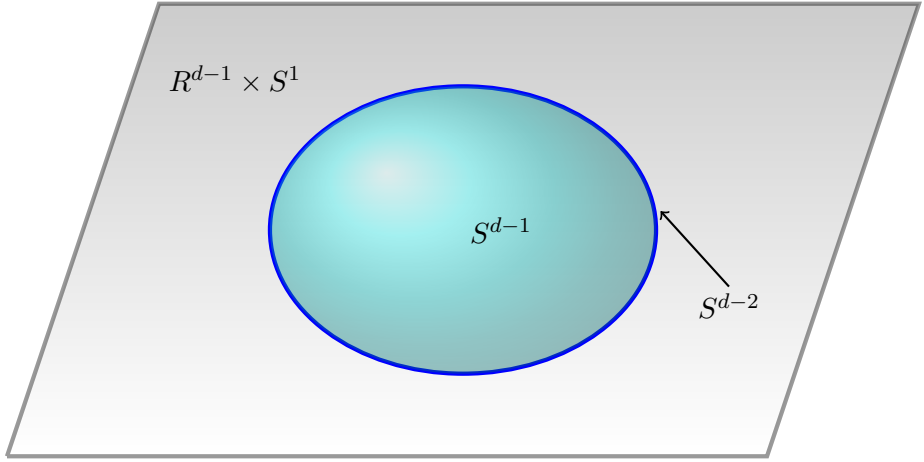

The action for a vector model marginally and conformally coupled on \(S^d\) is given by\cite{klebanov}\cite{romatschke} as: 

\begin{equation}\label{eq16}
   A^{(d)} =\frac{1}{2} \int_{s^d} dx \sqrt{-g}\bigg[\nabla_\mu \vec{\phi}\nabla^\mu \vec{\phi} + \frac{R(d-2)}{4(d-1)}\vec{\phi}^2+ \Big(\frac{\lambda}{N}\Big)^{\frac{2}{d-2}}\big(\vec{\phi}\cdot\vec{\phi}\big)^{\frac{d}{d-2}}\bigg],
\end{equation}
where \(g \equiv \det g_{\nu \mu}\) is the determinant of the metric tensor of \(S^d\), \(R\) is the Ricci scalar term defined as \( R = \frac{d(d-1)}{r^2}\) with \(r\) being the radius of the d-sphere. Again, the mass has been tuned to zero, resulting in a critical O(N) theory with conformal invariance on an odd d-dimensional Euclidean manifold under all couplings \(\lambda\). Further the entanglement entropy of a spherical region in \(S^{d-1}\) embedded in \(R^{d-1}\times S^1\) lives on the boundary in \(S^{d-2}\), as shown in figure (\ref{fig: entanglement entropy}), and maps directly to the thermal entropy of a theory on \(S^d\) \cite{klebanov}\cite{Myers}.

I will refer to the vacuum entanglement entropy of a region \(S^{d-1}\) which lives on \(S^{d-2}\), as \(s_{\text{EE}}^{(d)}\), and
\begin{equation}\label{eq17}
s_{\text{EE}}^{(d)} = \log |Z^{(d)}| =-\frac{1}{2} \log \big( \det{ \mathcal{O}^{(d)} }\big),
\end{equation}
\cite{klebanov} \cite{casini}, with, \(\mathcal{O}^{(d)} \equiv - \nabla^2 + \frac{R(d-2)}{4(d-1)}\), and \(Z^{(d)}\) being the partition function on \(S^{(d)}\). Eigenvalues of minus the Laplacian on a d-sphere for \(d\geq 2\) are given by \cite{Shimakura}\cite{klebanov} as
\begin{equation}\label{eq18}
    \lambda_n = n(n+d-1)r^{-2},
\end{equation}
with multiplicities, 
\begin{equation}\label{eq19}
 m_n=\frac{(2n+d-1)(n+d-2)!}{(d-1)!n!},\quad n\geq0.
\end{equation}
In odd dimension \(Z^{(d)}\) has no dependence on \(r\) giving a vanishing \(\beta\)-function. Thus, for simplicity, the dependence of \(r\) in \(\mathcal{O}\) is manipulated by letting \(r=1\). However, with hindsight of the effective mass \(m\) (associated with the auxiliary field \(\xi\)) being in the argument of logarithmic functions, I will keep the dimensionful quantity of r associated with \(m\).

\subsection{Entanglement entropy for d=3}

\begin{figure}[h]
\centering
\includegraphics[width=.9\textwidth]{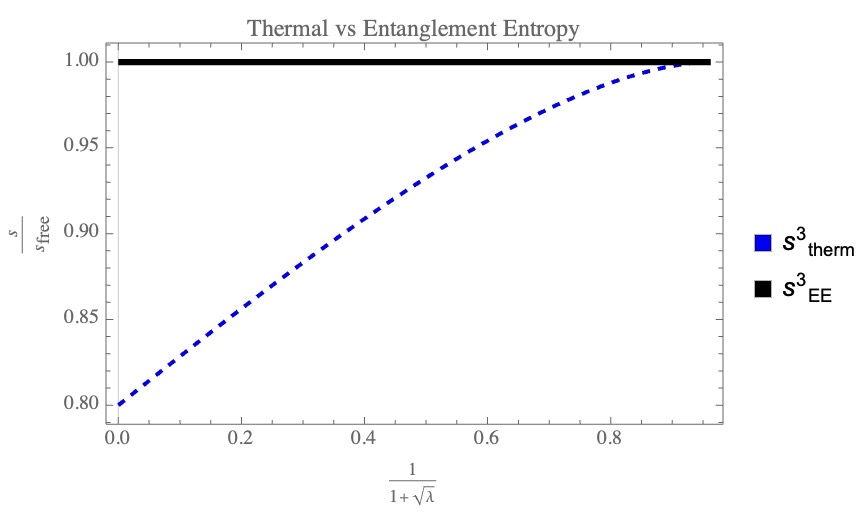}
\caption{\label{graph}For \(d=3\) the thermal entropy density of \(s_{\text{therm}}/s^{\text{free}}_{\text{therm}}\) is shown along with the entanglement entropy for a circular region that lives in \(R^2\times S^1\). Results are shown in the compactified domain of  \(\frac{1}{1+\sqrt{\lambda}} \in [0,1]\) starting at from the left at \(\lambda = \infty\).}
\label{fig: spheres}
\end{figure}

Considering \(d=3\) equation \eqref{eq3} can be modified so the integrals are now over \(S^3\) give three-dimensional hypersphere volumes. Then, 
\begin{equation}\label{eq20}
\begin{split}
s_{\text{EE}}^{(3)} = N\bigg[\frac{1}{2} \log \big( \det[ \mathcal{O} + (rm)^2] \big) - 4\pi r^3\bigg(\frac{2m^3}{9\sqrt{3}\lambda}\bigg) \bigg].
\end{split}
\end{equation}
Now the entanglement entropy of the free theory with the added auxiliary mass field is 
\begin{equation}\label{eq21}
s_{\text{EE}_{\text{free}}}^{(3)}=\frac{1}{2} \sum_{n=0}^\infty (n+1)^2 \log\big[ (n+1)^2 -\frac{1}{4} + (rm)^2\big].
\end{equation}
Taking the derivative of \(S_{\text{EE}_{\text{free}}}^{(3)}\) w.r.t. \(m^2\) and applying zeta-function regularization \cite{hawking} enables the saddle point of \(m\) to be evaluated: 
\begin{equation}\label{eq22}
\begin{split}
&\frac{\partial}{\partial m^2} s_{\text{EE}_{\text{free}}}^{(3)} \\&= \frac{1}{2} \sum_{n=0}^\infty\frac{(n+1)^2} {(n+1)^2 -\frac{1}{4} + (rm)^2}, \\&= \frac{1}{2}\bigg[\zeta(0) + \frac{1}{4}(2 - \sqrt{1 - (2rm)^2}\pi \cot\big(\frac{1}{2}\sqrt{1 - (2rm)^2}\pi\big)\bigg].
\end{split}
\end{equation}
Finally, utilizing equation \eqref{eq20} the saddle point condition for \(m\) is
\begin{equation}\label{eq23}
\frac{1}{8}\sqrt{1 - (2rm)^2}\pi \cot\big(\frac{1}{2}\sqrt{1 - (2rm)^2}\pi\big) - \frac{4\pi r^3m}{3\sqrt{3}\lambda} = 0.
\end{equation}
Equation \eqref{eq23} is satisfied for all couplings \(\lambda\) when \(m=0\). In evaluating \eqref{eq23}, the saddle condition of \(m\), call it \(C\big(m(\lambda)\big)\), is a functional of \(m\) which is a function of \(\lambda\), and \(C(m=0) = 0\) for all \(\lambda\). Therefore, \(m(\lambda)\) cannot be inverted into \(\lambda(m)\). Thus, the in-medium mass is vanishing, \(s_{\text{EE}}^{(3)}\) is invariant to coupling variations in the large \(N\) limit, and \(s_{\text{EE}_{\text{free}}}^{(3)}/ s_{\text{EE}_{\infty}}^{(3)} = 1\). Further, note satisfying \eqref{eq23} at \(m=0\) is effectively the same as setting the free propagator of the theory to zero. Recovering a functional form of equation \eqref{eq20} is achieved by; integrating equation \eqref{eq22}, evaluating for \(m=0\), and finally keeping the real parts of the result as demanded by the physical nature of the entanglement entropy. Likewise, removing the pure imaginary of integral result of \eqref{eq22} can be seen as removing an unspecified integration constant, then, 

\begin{equation}\label{eq24}
s_{\text{EE}}^{(3)}=-\frac{N}{16}\bigg(2\log(2) - \frac{3\zeta(3)}{\pi^2}\bigg).
\end{equation}
The entanglement entropy is negative, however, this is since in regularization we dropped the positively divergent non-universal contribution. This result matches N copies of the thermal entropy of a free massless theory on \(S^{3}\) given by \cite{klebanov}. It has further been shown by Ref. \cite{mezei}\cite{allais} that the d-sphere entanglement entropy represents a minimum among all shapes. Thus, the entanglement entropy on \(S^3\) in the large \(N\) limit is constant w.r.t. all couplings \(\lambda\) and minimized for all conformal geometries. Thus, while the thermal entropy in \(2+1d\) decreases monotonically with \(\lambda\), the entanglement entropy on a circular boundary in \(2+1d\) is strictly constant as shown in figure (\ref{fig: spheres}). 

\subsection{Generalizations to arbitrary d}

In order to generalize our theory to \(S^d\) first we can generalize the eigenvalues on \(S^d\), let them be \(\lambda^{(d)}_n\), and their associated multiplicities \(m^{(d)}_n\).
The eigenvalues take the form
\begin{equation}\label{eq25}
    \lambda^{(d)}_n = \Big(n + \frac{d-1}{2}\Big)^2 -\frac{1}{4} + (rm)^2,
\end{equation}
and the multiplicities generalize to, 
\begin{equation}\label{eq26}
    m^{(d)}_n = \frac{2}{(d-1)!}\prod_{j=0}^{\frac{d-3}{2}}\Big[\Big(n + \frac{d-1}{2}\Big)^2 - j^2\Big].
\end{equation}
Together equations \eqref{eq17}, \eqref{eq25}, and \eqref{eq26} give a general form for the saddle condition in \eqref{eq22} as
\begin{equation}\label{eq36}
    \frac{\partial}{\partial m^2} s^{(d)}_{\text{EE}} = \frac{-1}{(d-1)!}\sum_{n=0}^\infty \prod_{j=0}^{\frac{d-3}{2}} \frac{\big[(n + \frac{d-1}{2})^2 - j^2\big]}{(n + \frac{d-1}{2})^2 -\frac{1}{4} + (rm)^2} = 0.
\end{equation}
 In anticipation of the gap equation being solved by \(m=0\), the auxiliary field term can be ignored. This is effectively the same as working in the infinite coupling regime. In the case of \(d=5\), evaluating \eqref{eq36} at \(m=0\) gives 
\begin{equation}\label{eq28}
\begin{split}
    & \frac{\partial}{\partial m^2} S^{(5)}_{\text{EE}}\bigg|_{m=0}
    \\&= \frac{-1}{(4)!}\sum_{n=0}^\infty  \frac{(n + 2)^2\big[(n + 2)^2-1\big]}{(n + 2)^2 -\frac{1}{4} + (rm)^2}\bigg|_{m=0} \\&= \frac{-1}{(4)!}\big[\zeta(-2) -1 -\zeta(0)\frac{3}{4} + \frac{3}{4} -\frac{1}{8}] \\&= 0.
\end{split}    
\end{equation}
The integral of \eqref{eq28} evaluated at \(m=0\) further produces \(s^{(5)}_{\text{EE}}\) as
\begin{equation}
  s^{(5)}_{\text{EE}} =  \frac{N}{2^8} \bigg(2\log(2) + \frac{2\zeta(3) }{\pi^2} - \frac{15 \zeta(5)}
  {\pi^4}\bigg),
\end{equation}
matching N copies of the thermal entropy of a free theory in \(S^5\) given by Ref. \cite{klebanov}. 
This recipe can be repeated for \(d=7\) giving 

\begin{equation}
    s^{(7)}_{\text{EE}} = \frac{-N}{2^{12}}\bigg(4\log(2) + \frac{82\zeta(3)}{15\pi^2} - \frac{10\zeta(5)}{\pi^4} - \frac{63\zeta(7)}{\pi^6}\bigg)
\end{equation}

Note when evaluating \(s^{(d)}_{\text{EE}}\) for \(d=3\), \(d=5\), and \(d=7\) the process of zeta-function regularization produced a rational number which canceled when we set \(m=0\). For example, in equation \eqref{eq22} we have
\begin{equation}
\begin{split}
  &\frac{1}{2}\bigg[\zeta(0) + \frac{1}{4}(2 - \sqrt{1 - (2rm)^2}\pi \cot\big(\frac{1}{2}\sqrt{1 - (2rm)^2}\pi\big)\bigg]\Bigg|_{m=0} \\&= \frac{1}{2}\bigg[\zeta(0) + \frac{1}{4}(2)\bigg] = 0.
\end{split}
\end{equation}
This was likewise the case for \(d=5\) and \(d=7\). It is reasonable to ask if the effective mass is vanishing in the saddle condition for all odd dimensions \(d\), which is equivalent to asking if all rational terms produced in zeta function regularization will always perfectly cancel when setting \(m=0\). To explore this, modifying equation \eqref{eq36}, the derivative of the \(s^d_{\text{EE}}\) for arbitrary \(d\) can be expressed as 
\begin{equation}
\frac{\partial}{\partial m^2}\big(S^{(d)}_{\text{EE}}\big)=
-\frac{1}{2}\sum_{n=0}^{\infty}\frac{\Gamma(n+ d -1 )}{\Gamma(d) \Gamma(1 + n)}\frac{2 n+d-1}{
 n(n+d-1) + d/2(d/2-1) + (rm)^2},
\end{equation}
which evaluates analytically to 
\begin{equation}
\begin{split}
&\frac{\partial}{\partial m^2}\big(S^{(d)}_{\text{EE}})=-\frac{8\cos(d \pi/2)\cos(1/2 \sqrt{1 - 4 (rm)^2}\pi)}{\pi(d(d-2)+4(rm)^2))}\Gamma(1 - d)\\&\times\Gamma[1/2 (1 + d - \sqrt{1 - 4 (rm)^2})]\Gamma[1/2 (1 + d + \sqrt{1 - 4 (rm)^2})].
\end{split}
\end{equation} 
 Prior to regularization, it can be shown that for all odd positive integer values of \(d\) the saddle condition of \eqref{eq26} is satisfied in the limit \(m\rightarrow 0\), and likewise in the limits where \(d\rightarrow \infty\) and \(m\rightarrow 0\);
\begin{equation}\label{eq27}
\begin{split}
&\lim_{m\rightarrow 0} \frac{\partial}{\partial m^2}\big(S^{(d)}_{\text{EE}}\big) = 0 \quad \forall d \in\{ 2n-1|n\in\mathcal{N}\}\\& \And\\&
\lim_{m\rightarrow 0 
}\frac{\partial}{\partial m^2}\big(S^\infty_{\text{EE}})= 0.
\end{split}
\end{equation}

 Further, note that the integrating equation \eqref{eq23} and keeping the real parts gives a function that is monotonically decreasing for all \(m > 0\) and that all \(s^d_{\text{EE}}\) will likewise take the form of \eqref{eq23} before applying saddle conditions. Thus, there is only one global minimum for any \(s^d_{\text{EE}}\), and together with \eqref{eq27} this implies \(m=0\) is the only solution to the saddle point condition for odd \(d\) on \(S^d\). This also implies that all rational terms generated via zeta-function regularization will perfectly cancel for all odd \(d\), such as in the cases of \(d=3\), \(d=5\), and \(d=7\). Thus, via iteratively applying zeta-function regularization the entanglement entropy can be expressed as the real part of the integral form:
\begin{equation}\label{eq35}
\begin{split}
    &s^{(d)}_{\text{EE}}\\& = \lim_{m\rightarrow 0} \frac{N}{(d-1)!} (-1)^{(\frac{d+1}{2})} \Bigg[\int \prod_{j=0}^{\frac{d-3}{2}}\big[(rm)^2-\frac{1}{4}+j^2\big]\sum_{n=1}^\infty\frac{1}{n^2 -\frac{1}{4} +(rm)^2}dm^2 - C\Bigg],
\end{split}    
\end{equation}
where \(C\) is simply the rational term that is generated from the integral which will cancel with all rational terms generated via zeta-function regularization. Altogether, taking the limit and the integral in equation \eqref{eq35}, and then removing the rational and imaginary pieces will generate \(s^d_{\text{EE}}\) values that match the list of F-coefficients for free theories on \(S^d\) in Ref.\cite{klebanov} up a factor of \(-N\). Further, in this form, it can be shown that the entanglement entropy goes to zero in the limit \(d \rightarrow \infty\) with letting \(C=0\). 

\section{Conclusions}

In section 2) I derived previously known results for marginal coupling in the large N limit on \(R^2\times S^1\), via a generalized method which was then applied to higher dimensions. This resulted in a large class of CFTs, with non-polynomial potentials, which are solvable and monotonically decreasing for all couplings \(\lambda\) in \(d=3+4n\). In section \textbf{3} I further extended large \(N\) methods to \(S^3\) and showed the vacuum entanglement entropy \(s^{(3)}_{\text{EE}}\) is both solvable and constant for all couplings. This was then shown to hold for all \(s^{(d)}_{\text{EE}}\) given odd \(d\). The results of \(s^{(3)}_{\text{EE}}\) could be experimentally verified given a Bose-gas with a vanishing effective mass in 2+1 dimensions. Further, O(N) models are conjectured to have a high spin gravity dual by ref. \cite{klebanov2002}. Thus this work introduces the possibility of arbitrary even dimensional high spin gravity models which are solvable for all couplings.

Generally, this work prompts further questions on how coupling behaves on other novel manifolds in the Large N limit. Likewise, what are the physical motivations of this coupling invariant behavior, and is it unique to the hypersphere or is it found to exist on other manifolds more generally such as on a hyperbolic geometry? In Refs. \cite{allais} and \cite{mezei} perturbative and computational methods were used to generate entanglement entropy via small deformations of a sphere, however, no computational or analytic method has been provided to calculate the entanglement entropy of generally deformed spheres. Such as one can imagine deforming a sphere into flat space by letting the major axis of the deformed sphere go to infinity. Such methods could potentially conformally map the flat to the spherical theories presented in this paper, and further, bolster a general theory between field theory geometry and entanglement entropy. Further, this investigation of scalar theories could be extended to consider supersymmetric behavior in a large N Wess-Zumino model in \(S^d\) via methods similar to those used in \cite{dewolfe}, or consider numerical calculations of a \(\phi^4\) theory on \(S^3\) to better understand the conformal coupling invariance phenomena. In the extension to Fermionic theories, if it likewise holds that the effective mass is vanishing for all odd dimensional theories on \(S^d\), graphene could be used to experimentally test these theories. Likewise, one could extend this generalization of large N models to theories with even dimensions where the \(\beta\)-function is non-vanishing, and perhaps, likewise extend this large class of fully solvable models with non-polynomial potentials to even dimensional theories.

\acknowledgments

Many thanks to Paul Romatschke for the wonderful experience gained from working on this project together, and to Alex Flournoy for encouraging and supporting my independent study of QFT. This work was partially supported by the DOE, award number DE-SC0017905.  

\paragraph{Note added.}


\end{document}